\documentclass{elsart3p}
\journal{Physics Letters B}

\usepackage{hyperref}
\usepackage{amsfonts,amssymb,amsmath}
\usepackage{graphicx}
\usepackage[utf8]{inputenc}

\let\orgautoref\autoref
\renewcommand{\autoref}[1]{\def\equationautorefname{eq.}(\orgautoref{#1})}

\newcommand{\avg}[1]{\left\langle #1 \right\rangle }
\newcommand{\nucl}[2]{{}^{#2}\mathrm{#1}}

\begin{document}

\begin{frontmatter}



\title{Complete inclusion of parity-dependent level densities in the statistical description of astrophysical reaction rates}

\author[gsi,tu]{H. P. Loens},
\ead{h.p.loens@gsi.de}
\author[gsi,tu]{K. Langanke},
\author[gsi]{G. Mart\'inez-Pinedo},
\author[basel]{T. Rauscher} and
\author[basel]{F.-K. Thielemann}
\address[gsi]{Gesellschaft f{\"u}r Schwerionenforschung, Planckstr. 1, 64291 Darmstadt (Germany)}
\address[tu]{Technische Univerist\"at Darmstadt, Institut f\"ur Kernphysik, Schlossgartenstr. 9, 64289 Darmstadt (Germany) }
\address[basel]{Universit{\"a}t Basel, Klingelbergstr. 82, 4056 Basel (Switzerland)}

%!!! PREPRINT !!!


\begin{abstract}
Microscopic calculations show a strong parity dependence of the nuclear level density at low excitation energy of a nucleus. Previously, this dependence has either been neglected or only implemented in the initial and final channels of Hauser-Feshbach calculations. We present an indirect way to account for a full parity dependence in all steps of a reaction, including the one of the compound nucleus formed in a reaction. To illustrate the impact on astrophysical reaction rates, we present rates for neutron captures in isotopic chains of Ni and Sn. Comparing with the standard assumption of equipartition of both parities, we find noticeable differences in the energy regime of astrophysical interest caused by the parity dependence of the nuclear level density found in the compound nucleus even at sizeable excitation energies.
\end{abstract}

\begin{keyword}
parity dependence \sep Hauser-Feshbach theory \sep nuclear level density \sep neutron capture \sep astrophysical reaction rates \sep nucleosynthesis

\PACS 24.60.Dr \sep 21.10.Ma \sep 25.40.Lw \sep 26.30.-k
\end{keyword}
\end{frontmatter}

% main text
\section{Introduction}
Nuclear reactions in systems with high level density at low and intermediate energies
are commonly treated in the compound mechanism \cite{Gadioli:1992,Cowan:1991,Rauscher:2000}.
This reaction mechanism was
first postulated by Bohr in his well-known independence hypothesis, stating that
reactions can proceed via formation of a compound nucleus and that the decay of
the compound nucleus is determined entirely by its energy, angular momentum,
and parity, and not by the way it was formed \cite{Bohr:1936}. 
%The first quantitative
%theory based on the Bohr hypothesis was the Weisskopf-Ewing theory derived
%for reactions to the continuum. The more versatile Hauser-Feshbach
%theory developed later includes angular momentum conservation and thus allows to
%calculate reactions to discrete states (for an overview of this development,
%see e.g. \cite{Gadioli:1992}). It is the model of choice for
%many applications nowadays, including nuclear astrophysics.
%The fundamental
%Bohr 
This hypothesis remains valid below a projectile energy of a few tens of MeV.
At higher energies, doorway states, pre-compound, and direct reactions
become increasingly important. In the following, we focus on the low energy
region and thus on the pure statistical picture 
(Hauser-Feshbach theory, e.g. \cite{Gadioli:1992,Cowan:1991})  because our ultimate goal is
to propose an improved description for nuclear astrophysics. In astrophysical
nuclear burning, the relevant energy in the projectile-target system
does not exceed about 200
keV for neutron-induced reactions and $10$--$12$ MeV for proton- and $\alpha$-induced
reactions \cite{Rauscher:1997}. Most astrophysically important reactions
occur at even significantly lower energy.

Angular momentum conservation is included
in standard Hauser-Feshbach theory and thus
the Bohr hypothesis independently
holds for each spin $J$ and parity $\pi$
of the compound nucleus, formed from the interaction of
a projectile with a target nucleus.
There are two fundamental assumptions in the derivation of this theory:
1) There are always sufficient compound-nuclear states with $J^\pi$
in the relevant excitation energy range; 
2) the wave functions of the compound nuclear states have
random phases, so that interferences between reactions
proceeding through different compound nuclear states vanish.
Due to the strong energy-dependence of the nuclear level density, these
assumptions are valid for most reactions (especially at intermediate
energy) on stable targets studied in the laboratory. However, it
was shown \cite{Rauscher:1997} that the level density becomes too
low for the application of the Hauser-Feshbach statistical model
for astrophysically important reactions involving nuclei 
far off stability, exhibiting small particle
separation energies, or even for nuclei close to stability around
closed shells \cite{Descouvement:2006} at the low end of astrophysically
relevant energies.

For what follows it is important that in applications of the
Hauser-Feshbach model one assumes 
that both parities are equally present in the compound
nucleus at the formation energy. This presumption clearly is not
valid at very low excitation
energy (e.g. due to pairing effects), but there are also many indications from theory
as well as some experiments that parities may not be equilibrated
even at considerably large excitation energies, in some cases
up to 12 MeV \cite{Nakada:1997,Alhassid:2000,Mocelj:2007}.
Similar results are found in different approaches, e.g.,
in combinatorial methods using single-particle
energies from  microscopic Hartree-Fock-Bogoliubov
calculations \cite{Hilaire:2006}
as well as in recent Shell Model Monte Carlo (SMMC) calculations
\cite{Alhassid:2003,Ozen:2007}.
We note, however, that very recent  
experimental data for $^{58}$Ni and $^{90}$Zr are in accord with 
the equal-parity assumption at excitation energies as low as 7 MeV (for
$^{58}$Ni in disagreement with the theoretical predictions). Nevertheless,
the assumption of parity-independence of the level density
is clearly doubtful for a large number of exotic
nuclei in the energy range important in astrophysical environments.

In this Letter we incorporate the parity dependence of the level density
into the statistical description  of astrophysically capture reactions
in all stages of the
reaction.
Parity-dependent level densities have been
used before in statistical model calculations
\cite{Hilaire:2006,Mocelj:2007}. But these concern the
distribution of initial and final states, where the first can be populated due
to the finite temperature in the astrophysical environment,
while the
parity-dependence in the newly formed, excited compound
nucleus has never been considered before.
%because it does
%not appear in the regular Hauser-Feshbach equations due to
%the implicit assumptions made. 
Speaking in Bohr's terms, our
modification impacts the formation cross section of the compound
state.

In the next section we present the details of the modification.
This is followed by some restricted examples for application
to astrophysical neutron capture which are merely given to
discuss the model and to illustrate the possible implication for astrophysics.
The final section gives a summary and an outlook to future work.

\section{Formalism}

\begin{figure}
 \includegraphics[width=\columnwidth]{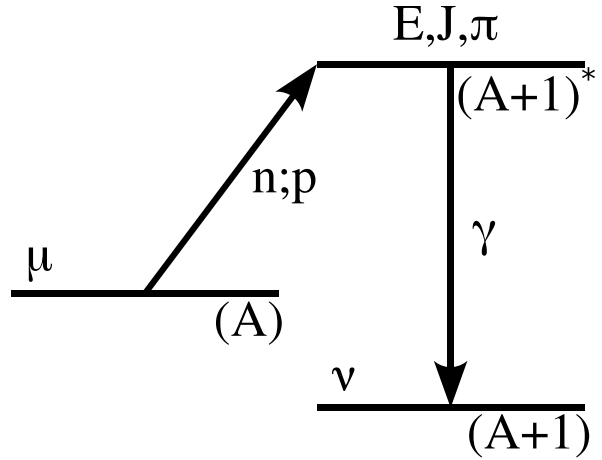}
 % schematic use of parity.eps: -1231081104x25 pixel, 0dpi, 659091923927040.00x-13384413.00 cm, bb=
 \caption{Schematic sketch of the compound capture reaction. A particle
 (neutron or proton) is captured in the state $\mu$ in the nucleus with mass
 number $A$, exciting a state in the compound/daughter nucleus $(A+1)$ at energy
 $E$ and with angular momentum and parity quantum numbers $J$ and $\pi$, 
 respectively. This state decays by $\gamma$ emission to the state $\nu$
 in the same nucleus.}
 \label{scheme}
\end{figure}

The Hauser-Feshbach expression for the cross section of a 
reaction
proceeding from the target state $\mu$ with spin $J_i^\mu$ and
parity $\pi_i^\mu$ to a
final state $\nu$ with spin $J_m^\nu$ and parity $\pi_m^\nu$
in the residual nucleus via a compound state with excitation
energy $E$, spin $J$, and parity $\pi$ (see Fig. \ref{scheme}) is given by
\begin{align}
 \sigma ^{\mu \nu}(E_{ij}) & = \dfrac{\pi \hbar^2}{2M_{ij}E_{ij}} \dfrac{1}{(2J^{\mu} _i + 1)(2J_j+1)} \notag \\ 
& \times \sum _{J,\pi} 
%\beta(E,\pi;J) 
\dfrac{T^{\mu}_j T^{\nu}
_o}{T_{\text{tot}}},
\label{Hauser}
\end{align}
where $E_{ij}, M_{ij}$ are the center-of-mass energy and the reduced mass, 
respectively, in the initial system,
while $J_j$ is the ground state spin of the projectile.
%and $\beta$ denotes
%a modification factor which we define below.
The transmission coefficients $T_j^\mu=T(E,J,\pi;E_i^\mu,J_i^\mu,\pi_i^\mu)$, $T_m^\nu=T(E,J,\pi;E_m^\nu,J_m^\nu,\pi_m^\nu)$ describe the transitions from the
compound state to the initial and final state, respectively. The sum of the
transmission coefficients of all possible channels are 
given by $T_\mathrm{tot}$.

%The transmission coefficients can be calculated by solving the 
%Schr\"odinger equation employing an appropriate optical potential.
As indicated, the transmission coefficients depend on the energy $E$ and the 
quantum numbers for angular momentum $J$ and parity $\pi$ of the
states excited in the compound nucleus. 
They can in principle be calculated by solving the Schr\"odinger equation
for the appropriate degrees of freedom. Such a microscopic approach will
describe simultaneously and consistently all the states in the compound nucleus
including their dependence on parity. Again in principle, this microscopic
Schr\"odinger equation can be mapped onto a complicated optical potential
which depends on energy and the other quantum numbers. There have been first
attempts to derive such microscopic potentials, which show indeed a strong
dependence on parity
\cite{Wintgen:1983,Langanke:1986}. However, these attempts are not 
realistic enough and are restricted to a few scattering systems. Thus in 
astrophysical (and other) applications of the Hauser-Feshbach model
the transmission coefficients are calculated
from optical potentials which are expected to give a reasonable
and global account for the many nuclei needed in nucleosynthesis 
calculations. For our discussion it is relevant
that these global optical potentials do not depend on parity and hence also
a possible parity dependence of the transmission coefficients is lost.
To overcome this shortcoming we propose here an indirect way.
It is based on the observation, that
for the average transmission coefficients, there is the 
relation $T \propto \avg{\Gamma}/D$,
involving the level spacing $D=1/\rho$ and the average level width
$\avg{\Gamma}$ in the considered reaction channel.
This linear proportionality
between transmission coefficient and level density $\rho$
leads us to define 
\begin{equation}
T(E,J,\pi)  = \beta(E,J,\pi) \hat{T}(E,J),
\label{modification}
\end{equation}
where $\hat{T}(E,J)$ is a transmission coefficient calculated for
a global, parity-independent potential (including centrifugal potential)
and the parity dependence is introduced by the weighting factor 
($\pi=\pm$)
\begin{equation}
\beta(E,J,\pi)  = 2 \cdot \frac{\rho(E,J,\pi)}{\rho(E,J,+) + \rho(E,J,-)}.
\label{Fullparity 3}
\end{equation}
The factor 2 accounts for the proper normalization. This approach assumes that the
parity dependence of the microscopic potential can be fully mapped onto the
level density appearing in the standard Hauser-Feshbach equations.
In the following applications we will use the same ansatz Eq. \ref{modification}
also for the transmission coefficients in the final channel.
As all transmission coefficients are evaluated at the same energy in
Eq. \ref{Hauser}, the $\beta$ factors of the total transmission
coefficient (denominator) and one of the $\beta$ factors of the nominator cancel.

As mentioned above, the Hauser-Feshbach approach assumes a ``sufficient''
number of levels in the excited compound nucleus so that an averaged
transmission coefficient $T$ is justified and the model is applicable. For
astrophysical application in the determination of astrophysical
reaction rates the incident energy distribution is given by a Maxwell-Boltzmann
distribution giving rise to a relevant
energy window \cite{Rauscher:1997,Rauscher:2000}. It has been shown that about
10 contributing levels (depending on which partial waves are dominating)
within this energy window are sufficient. This basic conclusion is not
affected by our treatment. However, the parity dependence may enhance or
reduce the number of available relevant levels and thus the applicability
limits have to be reevaluated taking into account the spins and parities of
the initial states. It should be noted that we do not change the total
level density but just distribute it differently between the parities.

%The transmission coefficients are calculated by solving the
%Schr\"odinger equation employing an optical potential. The optical potentials
%usually used are almost exclusively spin and parity independent. It has to
%be noted that microscopically derived optical potentials show a strong parity
%dependence \cite{Wintgen:1982}. Nevertheless, since such potentials are not available as global
%potentials we neglect this additional complication and utilize commonly used
%global potentials for the following examples.

\section{Results}
To explore the possible effects of our modification we
have performed a series of neutron capture cross section calculations.
At first we have performed conventional calculations in which we assumed parity equipartition
at all energies in target, compound nucleus and residual (calculation a).
Secondly we have restricted
the parity dependence to the level densities of the target and
residual nucleus (calculation b - these are similar to those of Mocelj {\it et
al.} \cite{Mocelj:2007}). Thirdly, we used a parity-dependent level density
in all three steps of the statistical treatment: the target, the
compound nucleus and the residual nucleus (calculation c).
Our calculations have been performed using
the spin and parity dependent level densities
of Hilaire and Goriely \cite{Hilaire:2006}. To explore how sensitive the results
depend on the set of level densities adopted we have repeated our
calculations using backshifted Fermi gas level densities with the
parametrizations as derived by Rauscher and Thielemann \cite{Rauscher:1997}
and the 
parity dependence as defined by Mocelj {\it et al.} \cite{Mocelj:2007}.

In this paper we focus on $(n,\gamma)$ for which the influence of the parity dependence of the level density can be discussed considering either the initial neutron capture or the final $\gamma$ decay. In the following we have chosen to consider the final $\gamma$ transitions which we assume to be either of parity-conserving
M1 or parity-changing E1 multipolarity. The reaction scheme is shown
in Fig. \ref{scheme}. Furthermore, we show astrophysical reaction
rates which include weighted sums over thermally excited states 
given by a thermal Maxwell-Boltzmann distribution
according to the conditions in a stellar plasma (for the relevant
definitions see, e.g., Ref.\ \cite{Rauscher:2000}).

As a first example 
we discuss the $^{58}$Ni(n,$\gamma$)$^{59}$Ni reaction for which
experimental data are available for comparison \cite{Dillmann:2006}.

\begin{figure}[!htb]
 \includegraphics[width=\columnwidth]{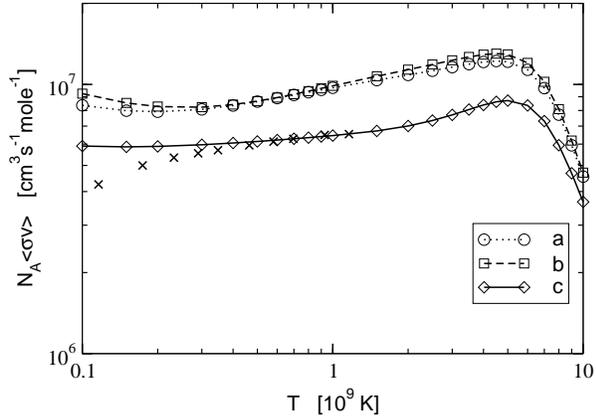}
 % ni58-paper.png: 1179666x1179666 pixel, 0dpi, infxinf cm, bb=
 \caption{Stellar reaction rate of 
$\nucl{Ni}{58}(n,\gamma)\nucl{Ni}{59}$; 
crosses: 
recommended values from 'KADoNIS v0.2' \cite{Dillmann:2006}, 
\textbf{(a)} without parity dependence, \textbf{(b)} using
parity dependence for the final states, \textbf{(c)} 
using parity dependence for the final states {\it and} the compound
formation using the level densities of \cite{Hilaire:2006}}
 \label{fig:ni58}
\end{figure}

\begin{figure}
 \includegraphics[width=\columnwidth]{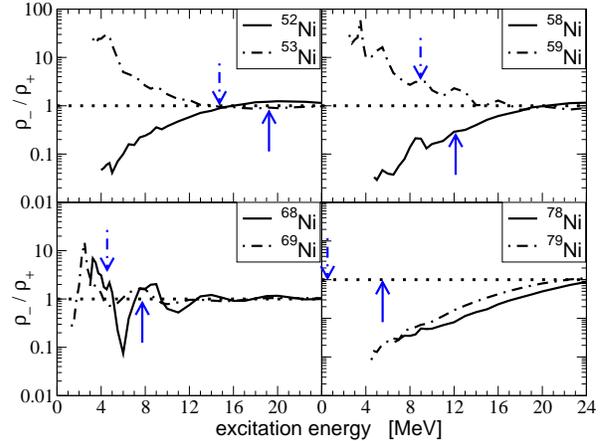}
  \caption{Ratios $\rho_{-}/\rho_{+}$ of several nickel nuclides; the parity distribution is from Hilaire \textit{et al.} \cite{Hilaire:2006} - we used $\rho(E,\pi)=\sum_J \rho(E,J,\pi)$ here; the arrows mark the neutron separation energy of the corresponding nickel isotope.}
 \label{fig:ratio}
\end{figure}

Fig. \ref{fig:ni58}
shows the  rate for this reaction 
as a function of temperature. One observes basically no difference between
calculations a) and b); i.e. the consideration of a parity dependence
in the level densities of the target and residual nucleus has no effect
in this case. This is in agreement with the findings of Mocelj {\it et al.}
\cite{Mocelj:2007}
and is mainly caused by the fact that both calculations use the 
experimentally known spectrum at low energies. However, considering
the parity dependence of the level density in the compound nucleus 
(calculation c) reduces
the rate by about $30\%$ which is a non-negligible effect.
The origin of this reduction becomes clear when one 
inspects Fig. \ref{fig:ratio}
which shows the ratio of parity-projected level densities 
for $\nucl{Ni}{58,59}$ defined as $\rho_{\pi} = \rho(E,\pi) = \sum_J \rho(E,J,\pi)$. We have summed over all spins as different values in $J$ have qualitatively the same dependence in the ratio $\rho_-/\rho_+$ plotted in Fig. \ref{fig:ratio}. For the
$(n,\gamma)$ reaction on $^{58}$Ni the $\gamma$ transitions in the
compound nucleus $^{59}$Ni are dominated by E1 multipolarity
at the relevant energies. At these
energies above the neutron threshold 
negative-parity states dominate the spectrum of $^{59}$Ni
due to the negative parity of the unpaired neutron occupying
single particle states in the
$pf$ shell. Thus initial states for E1 transitions into these states must have
positive parity and at stellar conditions they have to reside just above
the neutron threshold energy ($S_n=9$ MeV in $^{59}$Ni). At such
modest excitation energies the nuclear models predict still a dominance
of negative parity over positive parity states in the level density.
As we use the same {\it total} level density in all calculations a), b), and c),
the ratio $\rho_-/\rho_+ > 1$ at the relevant energies yields a reduction
of the dominant E1 transitions compared to a calculation which assumes
parity equipartition, i.e. $\rho_-/\rho_+ =1$. We note that this
reduction gets smaller with increasing temperature as then higher
excitation energies, at which the $\rho_-/\rho_+$ gets closer to unity,
contribute more to the stellar reaction rate.

For M1 transitions the effect is opposite as these require negative-parity
initial states for this reaction. These are enhanced 
compared to the parity-equipartition
assumption and hence the contribution of the M1 transitions 
relatively increase. This, however, has not much effect on the total
cross section which is dominated by E1 transitions.

It is also satisfying that our calculation yields a slightly better agreement
with the empirical data from the KADoNIS compilation when parity-dependent
level densities are incorporated into the statistical model (see Fig. \ref{fig:ni58}).

\begin{figure}[!htb]
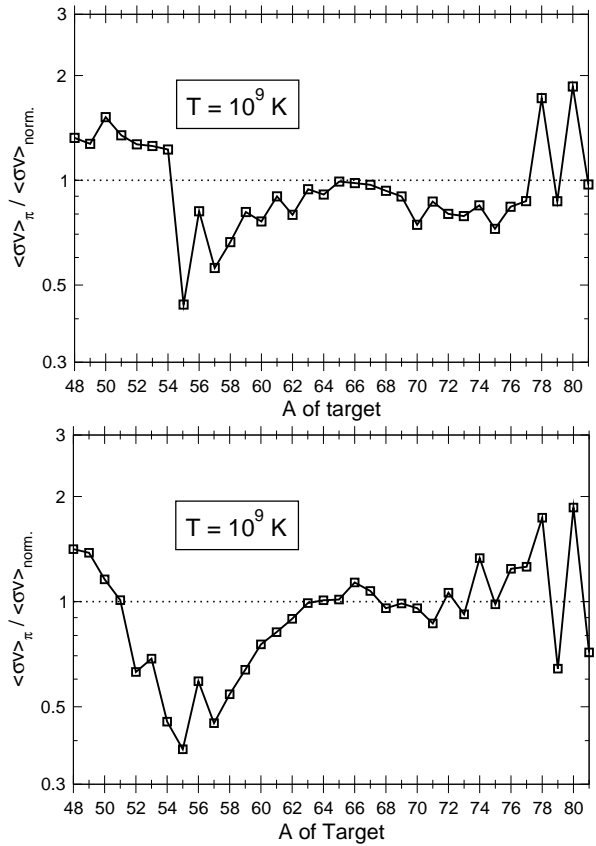

 \includegraphics[width=\columnwidth]{figure4a.eps}\\
 \includegraphics[width=\columnwidth]{figure4b.eps}
 \caption{Ratios of the stellar neutron capture rates of Ni isotopes with parity influence in target, compound, and residual to the stellar rate without any parity influence - the upper graph was made by using the level densities from \cite{Hilaire:2006} and the lower graph by using the parity distribution from \cite{Mocelj:2007} combined with a level density from \cite{Rauscher:1997}. The ratios are shown for a temperature $T=1$ GK of the stellar plasma.}
\label{fig:nichain}
\end{figure}

Fig. \ref{fig:nichain} shows the  neutron capture rates  for
the chain of nickel isotopes as obtained with full parity treatment
(calculation
c) relative to the standard treatment without parity dependence
(calculation a). To understand the
results one has to consider that
the importance of parity-dependent level densities  in
the statistical calculation of stellar reaction rates depends
on several ingredients:
1) the energy dependence of the ratio $\rho_-/\rho_+$,
2) the excitation energy of the neutron threshold in the compound,
3) the competition of parity-changing (E1) vs. parity-conserving
(M1) $\gamma$ transitions.

The ratio $\rho_-/\rho_+$ is shown in Fig. \ref{fig:ratio} for several
nickel isotopes. Obviously positive parity states dominate the low-energy
spectrum for even-even nuclei, while negative parity-states are more abundant
at low energies for odd-A nuclei. With increasing energy the ratio
$\rho_-/\rho_+$ tends to unity. However, the energy at which parity
equipartition is reached depends on nuclear structure, i.e. it is basically
determined by the energy difference of the Fermi energy to the nearest
level with different parity and the occupation of the levels.
Thus, the equipartition is achieved at increasingly higher
energies from $^{48}$Ni to $^{56}$Ni, where the $f_{7/2}$ and $d_{3/2}$
orbitals matter. By adding more neutrons it becomes 
energetically easier to excite those
to the $g_{9/2}$ level to make levels of opposite parity. As the energy
difference to this level decreases with increasing neutron number 
states with opposite parity (compared to the ground state parity) 
can be reached
at lower energies\footnote{The protons do not play an important role for 
determining the parity at low energies since nickel is a closed shell 
nucleus for protons.}. However, only the excitation of an odd number of
nucleons from or into the next oscillator 
shell changes the parity of the state. This leads to an oscillatory
behavior in the $\rho_-/\rho_+$ ratios at energies already below the
neutron thresholds (indicated by arrows in Fig. \ref{fig:ratio}). 
For $^{68}$Ni with the neutron number $N=40$
the $pf$ shell is completely occupied in the independent particle model.
Hence parity-changing transitions appear at quite low energies
in the nickel isotopes around $^{68}$Ni. Moving to even larger neutron numbers,
and thus closer to potential r-process nuclei around $^{78}$Ni, nickel isotopes
have mainly positive parity states at low energies
as orbitals in the $gds$ shell have positive parity. We note further that 
equipartition is reached at somewhat higher energies (about 3 MeV) in even-even
nuclei than in odd-A nuclei due to pairing.

Fig. \ref{fig:ratio} also shows the neutron separation energies
\cite{Audi:2003} which obviously decrease with increasing neutron 
number along an
isotope chain. The odd-even staggering is due to pairing.

Usually E1 transitions, which are modelled by a Lorentzian centred around
the giant dipole resonance in our approach
\cite{Cowan:1991}, dominate over M1 transitions, which 
are described by the single-particle model which makes the M1 strength
energy independent \cite{Cowan:1991}. However, if the capture
occurs at energies significantly below the giant dipole resonance, E1
transitions are strongly suppressed relative to M1 transitions in this model
and the latter can dominate. This can occur 
in very neutron-rich isotopes with very low neutron thresholds.

For the nickel isotopes with largest proton excess we calculate an enhancement
in the stellar rates if the parity dependence of the level densities
is considered (Fig. \ref{fig:nichain}). 
For these nuclei the neutron thresholds are quite high
and, due to the excitation of an odd number of nucleons from the $ds$ shell,
there is an enhancement of states in the compound 
with opposite parity to the ground state.
This yields a slight increase in the rate. The larger effect
stems from the fact that for these nuclei low-energy states are
experimentally not known and have to be modelled. These states, however,
are likely to have the same parity as the ground state 
(positive for even-even and negative
for odd-A proton-rich isotopes). Considering this fact increases
the rate compared to calculations which assume
parity equipartition also at low excitation energies.

For the 
isotopes $^{55-62}$Ni we observe a reduction of the rate,
if parity dependence is considered. The origin is the same as  
discussed above for the case of $^{58}$Ni, as E1 transitions are 
reduced as 
$\rho_+/\rho_- <1 $ ($\rho_-/\rho_+ < 1$)
at the energies just above the neutron threshold for odd-A (even-even)
isotopes. 

For the isotopes $^{63-72}$Ni the neutron thresholds are located at energies
where the ratio of parity-dependend level densities is rather close to unity,
but still shows some oscillatory behavior. As a consequence the rates
are slightly enhanced or reduced depending on the fact whether
the ratio is just above or below unity at the energies above the
neutron threshold. 

For the most neutron-rich Ni isotopes, 
states with the same parity as the ground state dominate the spectrum
at energies around the neutron threshold. This leads to a reduction
of the rate if E1 transitions dominate. However, for the capture
on the nickel isotopes $^{78}$Ni and $^{80}$Ni the thresholds are so low
(0.52 MeV and 0.17 MeV, respectively)
that M1 transitions contribute more in our model than E1 captures; hence
the rate is increased compared to the case where  parity equipartition
is assumed. Due to pairing the neutron threshold in $^{80}$Ni is at
3.7 MeV and E1 capture dominates. We note that these nuclei are
close to the r-process path.
Hence our discussion clearly shows that the effect of 
parity-dependend level densities on the neutron capture rate is quite
sensitive to the neutron separation energies and the competition of
M1 and E1 transitions which are both not sufficiently well known yet.
We also note that the low density of states makes the use of a statistical
model for the very neutron-rich isotopes questionable. Furthermore
direct neutron captures should contribute to the rates for these nuclei.
Here the parity dependence of the optical potential should be incorporated into
the models and possible effects studied.

We have repeated our calculation of neutron captures on the nickel isotopes
using the parity distribution of Mocelj \textit{et al.} 
\cite{Mocelj:2007} combined with the backshifted Fermi gas model of 
Rauscher {\it et al.} \cite{Rauscher:1997}. As is shown 
in the lower graph of Fig.
\ref{fig:nichain} the effects are quite similar to those obtained with the
Hilaire and Goriely level densities \cite{Hilaire:2006}. 

\begin{figure}[!htb]
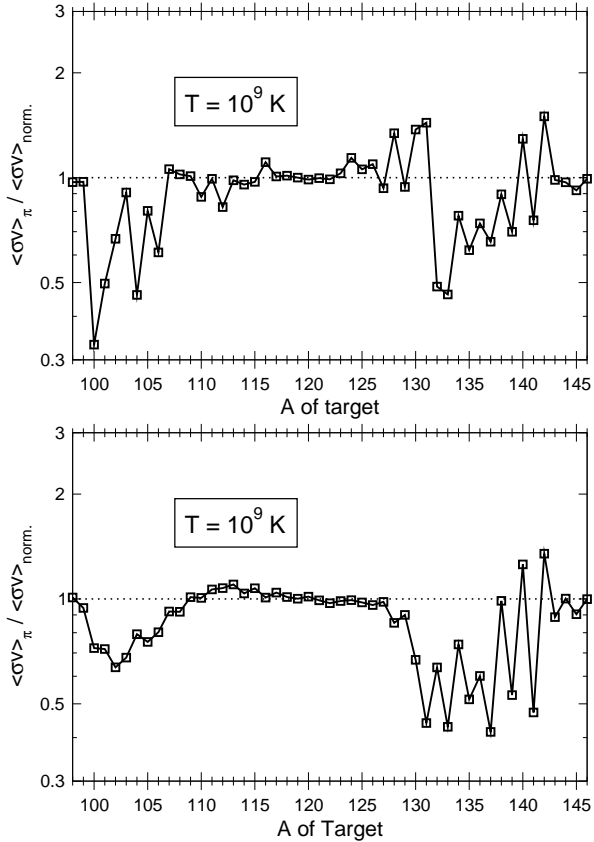

 \includegraphics[width=\columnwidth]{figure5a.eps}\\
 \includegraphics[width=\columnwidth]{figure5b.eps}
 \caption{Ratios of the stellar neutron capture rates of Sn isotopes with parity influence in target, compound, and residual to the stellar rate without any parity influence - the upper graph was made by using the level densities from \cite{Hilaire:2006} and the lower graph by using the parity distribution from \cite{Mocelj:2007} combined with a level density from \cite{Rauscher:1997}. The ratios are shown for a temperature $T=1$ GK of the stellar plasma.}
\label{fig:snchain}
\end{figure}

Finally we have performed a series of calculations for the
neutron capture on the tin isotopes (Fig. \ref{fig:snchain}). Again 
E1 transitions dominate and for the same reasons as explained in details
above (e.g. for the case of $^{58}$Ni) the consideration of 
parity-projected level densities lead to
a reduction of the rates. For the tin isotopes $^{120-130}$Sn the 
neutron intruder
state $h_{11/2}$, with the opposite parity to the other
orbitals in the $gds$ shell, plays a special role which has no equivalent
for the nickel isotopes. It leads to a rather fast parity equilibration
in the level densities which reaches 
ratios $\rho_-/\rho_+$ close to unity at energies around
the neutron thresholds \cite{Hilaire:2006}. 
As a consequence the rates for neutron captures on the mid-shell tin isotopes
change only mildly
if a parity dependence of the level densities
is considered. For the heavier tin isotopes, the $N=82$ neutron shell gap at
$^{132}$Sn and the fact that the two lowest single particle orbitals
beyond $N=82$ ($f_{7/2},p_{3/2}$) have the same parity 
as the $h_{11/2}$ intruder level
have the effect that the equipartition of parities is reached at
larger excitation energies than the respective neutron thresholds for 
tin isotopes beyond $^{132}$Sn. For 
similar reasons as for
the case of $^{58}$Ni, the capture rate is reduced for
these tin isotopes. The odd-even dependence in the rates are caused by
the pairing effects in the neutron threshold energies.
For the even heavier tin isotopes
the neutron intruder state $i_{13/2}$, with opposite parity to
$h_{11/2},f_{7/2},p_{3/2}$ becomes important resulting in a fast
parity equipartition of the level densities. As a consequence
the capture rates do not change much, if parity-dependent level 
densities are used.

\section{Conclusion}
We have presented a simple method in the framework of the statistical Hauser-Feshbach theory to account for a full parity dependence including non-uniformly distributed parities in the nuclear level density of the compound nucleus. This goes beyond previous work which only accounted for parity-dependent level densities in the initial and final channels but not in the compound step of the reaction. We applied our method to capture reactions on Ni and Sn nuclei, using a parity dependence in all steps of the compound nucleus reaction. We conclude that this treatment can have a noticeable effect on astrophysical reaction rates for nuclei far from stability. In principle, our approach can also be extended to include a spin dependence or a more general dependence on the level density of the compound nucleus. This will be the focus of future work.

This work was performed in the framework of the SFB634 of the Deutsche Forschungsgemeinschaft and supported by the Swiss National Science Foundation (grant 2000-105328).

% \appendix

% \section{}
% \label{}
%\bibliographystyle{nucphys_noURL}
%\bibliographystyle{apsrev}
%\bibliography{Biblio}

\end{document}